\documentclass[]{pasj02} 
\usepackage[switch,mathlines]{lineno} 
\usepackage{natbib} 

\jyear{2026}
\Received{2026/01/30}
\Accepted{2026/03/07}

\newcommand{\Ni}{\ensuremath{^{56}\mathrm{Ni}}\xspace}

\newcommand{\Msun}{\,\ensuremath{\mathrm{M}_\odot}\xspace}

\newcommand{\Msunpyr}{\,\ensuremath{\Msun~\mathrm{yr^{-1}}}\xspace}

\graphicspath{{./}{figures/}} 

\usepackage{xspace}

\begin{document} 

\title{ Supernovae interacting with Si and S-rich circumstellar matter from double white dwarf mergers }

\author{
 Takashi J. \textsc{Moriya},\altaffilmark{1,2,3}\altemailmark\orcid{0000-0003-1169-1954} \email{takashi.moriya@nao.ac.jp}
 Chengyuan \textsc{Wu},\altaffilmark{4}\orcid{0000-0002-2452-551X} 
 Dongdong \textsc{Liu},\altaffilmark{4}\orcid{0000-0002-3007-8197}
 Zheng-Wei \textsc{Liu},\altaffilmark{4}\orcid{0000-0002-7909-4171}
 and 
 Bo \textsc{Wang}\altaffilmark{4}\orcid{0000-0002-3231-1167} 
}
\altaffiltext{1}{National Astronomical Observatory of Japan, National Institutes of Natural Sciences, 2-21-1 Osawa, Mitaka, Tokyo 181-8588, Japan}
\altaffiltext{2}{Graduate Institute for Advanced Studies, SOKENDAI, 2-21-1 Osawa, Mitaka, Tokyo 181-8588, Japan}
\altaffiltext{3}{School of Physics and Astronomy, Monash University, Clayton, VIC 3800, Australia}
\altaffiltext{4}{International Center of Supernovae (ICESUN), Yunnan Key Laboratory of Supernova Research, Yunnan Observatories, Chinese Academy of Sciences, Kunming 650216, China}




\maketitle

\begin{abstract}
We present that supernovae interacting with a dense Si and S-rich circumstellar matter like SN~2021yfj can originate from mergers of two white dwarfs. A C+O white dwarf accreting He from its non-degenerate He companion star can initiate a C burning frame at its surface propagating inward under certain conditions. Such a burning frame synthesizes intermediate mass elements such as Si and S, forming a hybrid WD with an outer Si+S-rich layer. After the He star companion becomes a white dwarf, the two white dwarfs can eventually merge. During the merger, the outer layers of the hybrid white dwarf can be tidally stripped, forming a dense Si and S-rich circumstellar matter. If a thermonuclear explosion is triggered in the merging white dwarfs, an explosion within a dense Si and S-rich circumstellar matter can be realized, resulting in SN~2021yfj-like events. We argue that the properties of SN~2021yfj can be reproduced by a dense Si and S-rich circumstellar matter having $\simeq 0.3~\Msun$ within which an explosion having kinetic energy of $\simeq 4\times 10^{50}~\mathrm{erg}$ and ejecta mass of $\simeq 0.3~\Msun$ occurred. These properties are consistent with the double white dwarf merger scenario. This scenario can naturally explain the existence of He observed in SN~2021yfj. Because white dwarf mergers can also lead to the formation of He and C+O dense circumstellar matter, some Type~Ibn and Icn supernovae may also originate from a similar evolutionary path.
\end{abstract}


\section{Introduction}
Extensive transient surveys in the last decade unveiled tremendous diversities in stellar deaths \citep[e.g.,][]{perley2020}. Especially, many supernovae (SNe) are discovered to have circumstellar matter (CSM) much denser than predicted by stellar evolution theory, indicating the existence of unknown mass loss activities occurring shortly before the terminal stellar explosions (\citealt{dessart2024} for a recent review). The interaction between SN ejecta and dense CSM results in the formation of narrower lines than observed in typical SN spectra. SNe with narrow lines are noted by ``n'' in the SN classification scheme like Type~IIn SNe (SNe~IIn), indicating that they have narrow H emission line features \citep{schlegel1990}.

Core-collapse SNe are broadly classified into three spectroscopic classes: SNe with H (SNe~II), SNe without H but with He (SNe~Ib) and else (SNe~Ic which mostly show C+O features, \citealt{filippenko1997}). These spectroscopic classes are known to have the corresponding ``n'' types (SNe~IIn, Ibn, and Icn). SNe without H but with strong Si features are classified as SNe~Ia and they originate from thermonuclear explosions of white dwarfs (WDs, e.g., \citealt{nugent2011}). SNe~Ia sometimes show narrow H features, and such SNe~Ia are often called ``SNe~Ia-CSM'' \citep[e.g.,][]{silverman2013,sharma2023}. SNe~Ia are also found to show interaction signatures with dense He \citep{kool2023} and C+O \citep{tsalapatas2025} CSM, showing the diverse progenitor channels leading to SN~Ia explosions.

Recently, \citet{schulze2025} reported the discovery of SN~2021yfj, which shows strong narrow emission features of Si and S unveiling the existence of SNe interacting with Si and S-rich dense CSM. Interestingly, SN~2021yfj also showed narrow He emission features. The mechanism to form such a dense Si and S-rich CSM is not trivial. \citet{schulze2025} suggested that the progenitor of SN~2021yfj is a massive star. Massive stars form Si and S-rich layers as a result of O burning within several years before core collapse \citep{woosley2002}. If this Si and S-rich layer, as well as all the outer layers with lighter elements, can be somehow ejected before the core collapse to form a Si and S-rich CSM, the subsequent core-collapse explosion can result in interacting SNe with the Si and S-rich CSM. However, releasing the Si and S-rich layers located at the innermost layers of massive stars shortly before their core collapse is challenging. The existence of He is also difficult to explain in this massive star scenario because He should have been burnt at the innermost layers of massive stars. \citet{schulze2025} speculated that He may originate from a companion star.

In this paper, we present that SN~2021yfj-like SNe interacting with Si and S-rich CSM can originate from mergers of two WDs. As also briefly discussed in \citet{schulze2025}, He burning on the surface of a WD can result in the formation of Si and S. \citet{schulze2025} discarded this possibility by arguing that the formed Si and S are likely continue to be fused to heavier nuclei because of high temperature required for He burning. However, \citet{wang2017,wu2019} showed that He accretion onto a C+O WD can lead to a surface C ignition launching a inward burning frame forming Si and S-rich outer layers in the WD. The burning frame is suggested to quench at some moment, forming a hybrid WD having inner C+O layers and outer O+Si+S layers \citep{wu2020}. If the mass of such a hybrid WD becomes close to the Chandrasekhar limit and a thermonuclear explosion is triggered, it could be observed as a SN~Ia with high-velocity Si features due to the existence of Si in the outer layer of the hybrid WD \citep{wu2020}. Here, we propose that such a hybrid WD can result in SN~2021yfj-like events if it is in a double WD (DWD) system and experience a merger by gravitational wave energy loss like a so-called ``double-degenerate'' scenario for SNe~Ia \citep{iben1984,webbink1984}.

The rest of this paper is organized as follows. We present an example of the evolutionary pathway leading to SN~2021yfj-like events through DWD mergers in Section~\ref{sec:progenitor}. We overview the observational properties of SN~2021yfj and discuss if it is consistent with the DWD merger scienario in Section~\ref{sec:sn2021yfj}. We discuss our results and conclude this paper in Section~\ref{sec:discussion}.

\section{Progenitor evolution}\label{sec:progenitor}
\begin{figure}
 \begin{center}
  \includegraphics[width=9.3cm]{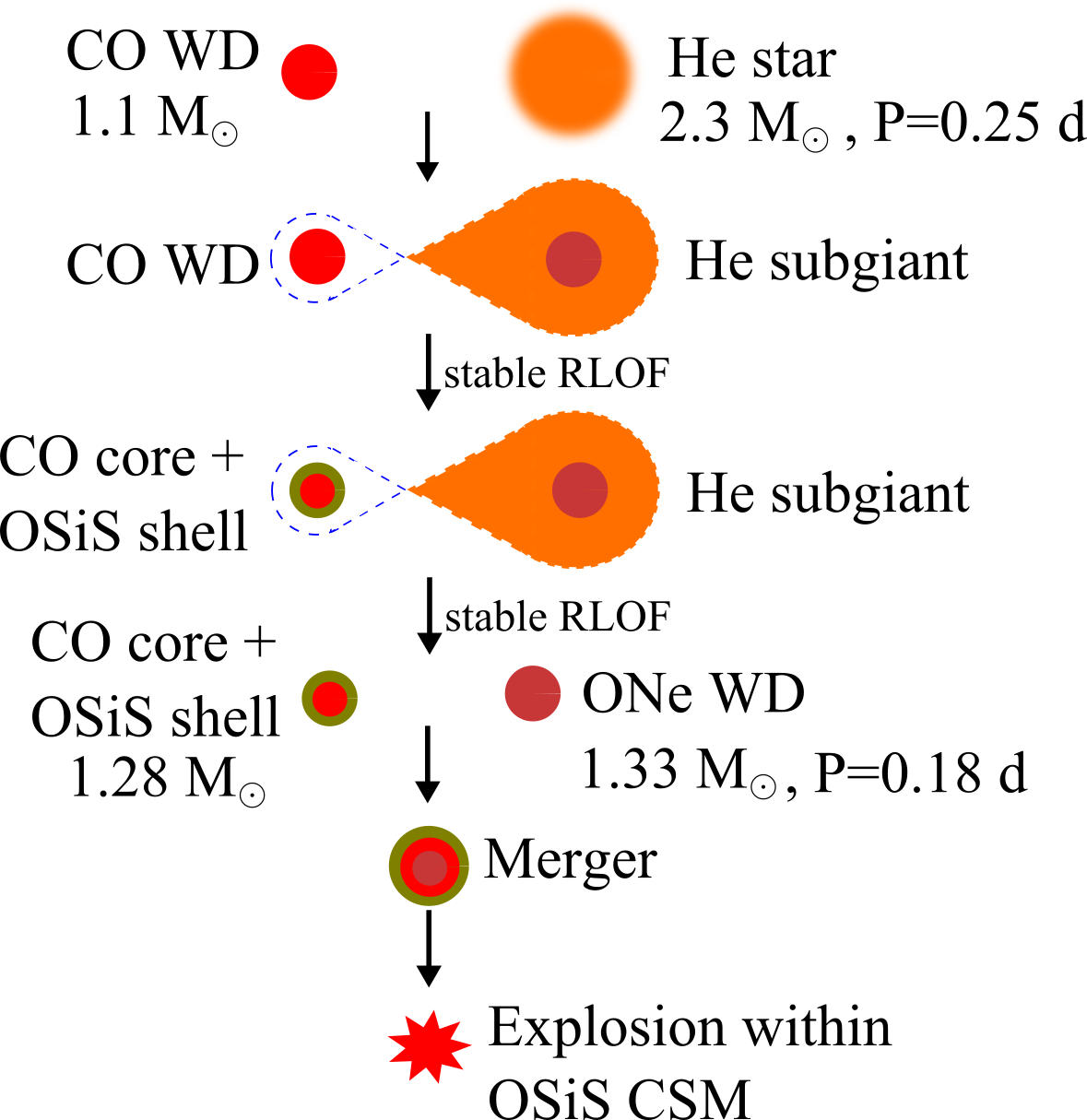} 
 \end{center}
\caption{ Schematic illustration of an evolutionary path leading to a DWD merger forming Si and S-rich CSM.
 {Alt text: A picture showing the evolutionary path of two stars. The evolution proceeds from top to bottom.} 
}\label{fig:illustration}
\end{figure}

Figure~\ref{fig:illustration} provides a schematic illustration of the evolutionary path leading to the formation of a Si and S-rich CSM within which an explosion can occur. This explosion can lead to SN~2021yfj-like events as explained in the next section. Here, we introduce the evolutionary path. An example of the initial mass and final WD masses, He star masses, and orbital periods obtained by binary population synthesis calculations by the same method as in \citet{liu2017,liu2018} are provided as a reference.

The initial configuration is a close binary system of a C+O WD and a He star. As the He star expands, stable Roche-lobe overflow (RLOF) starts and mass transfer from the He star to the C+O WD is initiated and He is accreted onto the WD. If the He accretion rate exceeds a critical value of around $2\times 10^{-6}~\Msunpyr$, off-center carbon burning can be ignited near the surface \citep{wang2017}. The carbon flame then propagates inward and eventually quenches \citep{wu2019,wu2020}, leading to the formation of a hybrid WD consisting of an inner C+O core and an outer layer enriched in O and intermediate-mass elements such as Si and S. An example the abundance profile of such a hybrid WD computed by MESA \citep{paxton2011,paxton2013,paxton2015,paxton2018,paxton2019,jermyn2023} with the method explained in \citet{wu2019,wu2019b,wu2020} by assuming a constant He accretion rate of $5\times{10}^{-6}~\Msunpyr$ is presented in Figure~\ref{fig:abundance}. The initial mass of the C+O WD in this computation is $1.1~\Msun$. It shows that the outer layers of the hybrid WD can naturally develop non-negligible abundances of Si and S, providing the required CSM composition for SN 2021yfj. A detailed presentation of the WD evolutionary models and their resulting abundance profiles will be given in a follow-up paper (C. Wu et al. in preparation).

The He star eventually loses all the He layer to be a WD and a DWD system is formed. The DWD system merges after losing the orbital energy through gravitational waves. During the merger, the outer layers of the less massive WD can be tidally stripped to form a CSM. If the hybrid WD is less massive, its outer layers are tidally stripped to form the CSM containing Si and S. Because the other WD needs to be more massive than the hybrid WD having $\simeq 1.3~\Msun$ to tidally strip the hybrid WD, the other WD from the He star is likely a massive O+Ne WD. C and O, which consist the inner layers of the tidally-stripped hybrid WD, are accreted onto the O+Ne WD. The outcome of the accretion is uncertain \citep[e.g.,][]{wu2023b}. However, it is possible that explosive C burning is ignited to trigger the thermonuclear explosion of the inner WD system \citep{dan2014,kashyap2018}. Then, an explosion under the Si and S-rich CSM leading to an interaction-powered SN like SN~2021yfj can be triggered. Depending on the ejecta mass and CSM mass, a bound remnant could be left behind. If the bound remnant is massive enough, it can eventually explode as an electron-capture SN in $100-1000$~years after the merger \citep{wu2023b}.

\section{SN~2021yfj}\label{sec:sn2021yfj}
In this section, we discuss if the DWD merger event discussed in Section~\ref{sec:progenitor} can explain the properties of SN~2021yfj. For this purpose, we first estimate the SN ejecta and CSM properties of SN~2021yfj. We assume that the rise time of interaction-powered SNe represents the diffusion time $t_\mathrm{d}$ in the CSM. Using the optical depth in the CSM $\tau_\mathrm{CSM}$ and the CSM radius $R_\mathrm{CSM}$, we can express $t_\mathrm{d}\simeq\tau_\mathrm{CSM}R_\mathrm{CSM}/c$, where $c$ is the speed of light. Motivated by the fact that SN~2020aeuh, a SN~Ia with a dense CSM likely formed by tidal stripping during a DWD merger \citep{tsalapatas2025}, likely had a flat CSM density structure, we assume that the CSM has a constant density $\rho_\mathrm{CSM}$. In this case, the diffusion time can be expressed as $t_\mathrm{d}\simeq \kappa\rho_\mathrm{CSM}R_\mathrm{CSM}^2/c$, where $\kappa$ is the CSM opacity and is assumed to be $0.1~\mathrm{cm^2~g^{-1}}$. The rise time of the bolometric light curve of SN~2021yfj was about 5~days\footnote{The time it took to reach the peak luminosity from the half of the peak luminosity is estimated to be 2.3~days \citep{schulze2025}. We set the rise time to be 5~days by assuming that a similar time was required to reach the half of the peak luminosity.} and the blackbody radius at the peak luminosity was $10^{15}~\mathrm{cm}$ \citep{schulze2025}. Thus, we set $t_\mathrm{d}=5~\mathrm{days}$ and $R_\mathrm{CSM}=10^{15}~\mathrm{cm}$. With these assumptions, we can estimate $\rho_\mathrm{CSM}=8\times10^{-14}~\mathrm{g~cm^{-3}}$. The corresponding CSM mass is $0.3~\Msun$. We note that the CSM is optically thick ($\tau_\mathrm{CSM}\simeq 8$).

For the purpose of an order of magnitude estimates, we assume that the peak bolometric luminosity of SN~2021yfj ($4\times 10^{43}~\mathrm{erg~s^{-1}}$) roughly corresponds to the luminosity input from the interaction at around the peak luminosity \citep{arnett1982,chatzopoulos2012}, i.e., $2\pi\varepsilon R_\mathrm{CSM}^2\rho_\mathrm{CSM}(v_\mathrm{shock}-v_\mathrm{CSM})^3$, where $\varepsilon$ is the conversion efficiency from kinetic energy to radiation energy, $v_\mathrm{shock}$ is the shock velocity, and $v_\mathrm{CSM}$ is the unshocked CSM velocity. Assuming $\varepsilon=0.3$, we can estimate that $v_\mathrm{shock}-v_\mathrm{CSM}\simeq 5000~\mathrm{km~s^{-1}}$. Assuming $v_\mathrm{CSM}\simeq 1000~\mathrm{km~s^{-1}}$, the shock velocity is $v_\mathrm{CSM}\simeq 6000~\mathrm{km~s^{-1}}$. This velocity is faster than the fastest velocity component observed in the spectra of SN~2021yfj ($3000~\mathrm{km~s^{-1}}$, although He may have higher velocity components). Because we intend to have an order of magnitude estimate estimate here, we argue that our simple model is roughly consistent with the observed properties of SN~2021yfj.

It is difficult to break the degeneracies between the explosion energy and ejecta mass. The ejecta mass from inside needs to be comparable or less than the CSM mass to efficiently decelerate the ejecta to be observed as an interaction-powered SN. If we assume the ejecta mass $M_\mathrm{ej}$ is similar to the CSM mass ($M_\mathrm{CSM}=0.3~\Msun$), the conservation of momentum, $M_\mathrm{ej}v_\mathrm{ej}+M_\mathrm{CSM}v_\mathrm{CSM}=(M_\mathrm{ej}+M_\mathrm{CSM})v_\mathrm{sh}$ where $v_\mathrm{ej}$ is the ejecta velocity, provides the SN ejecta energy $E_\mathrm{ej}$ estimate of $E_\mathrm{ej}=M_\mathrm{ej}v_\mathrm{ej}^2/2=4\times10^{50}~\mathrm{erg}$.
A small amount of \Ni, likely of the order of $0.01~\Msun$ at most \citep[e.g.,][]{zetani2023,kashyap2018}, may be produced during the merger but the \Ni\ mass is too small to affect the early light curve. The late-phase luminosity could have been affected by the radioactive decay, although it was not observed in SN~2021yfj.
We will investigate light-curve properties of the DWD merger models including the effects of the CSM interaction and the \Ni\ decay in detail in a follow-up paper (T.J. Moriya et al. in preparation).

\begin{figure}
 \begin{center}
  \includegraphics[width=8cm]{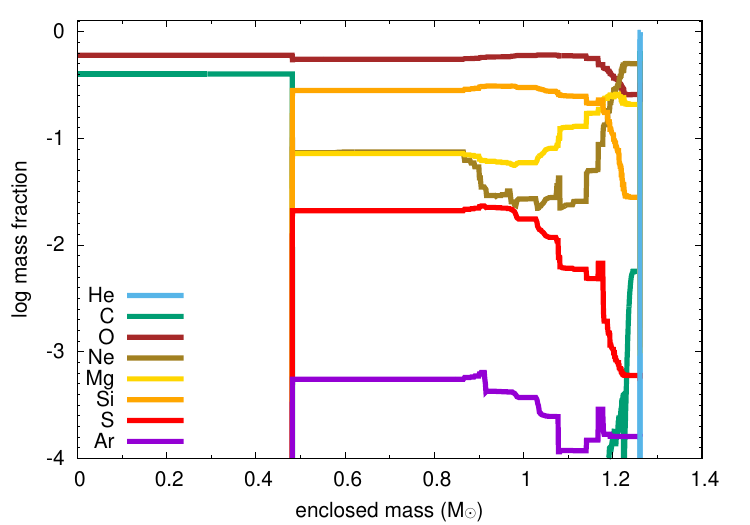} 
 \end{center}
\caption{ An example of the chemical composition of the hybrid WD formed through the He accretion initiating a C burning frame propagating inward.
 {Alt text: Lines show the mass fraction of each element at the mass coordinate.  The x axis is the mass coordinate from 0 to 1.4 solar masses. The y axis is the abundance with log ranging from -4 to 0.1. } 
}\label{fig:abundance}
\end{figure}

\begin{table}
  \tbl{Average mass fraction of elements in the outer $0.3~\Msun$ of the hybrid WD in Fig.~\ref{fig:abundance}.}{%
  \begin{tabular}{cc}
      \hline
      Element & Mass fraction  \\ 
      \hline
      He & 0.004  \\
      C  & 0.001 \\
      O  & 0.5  \\
      Ne & 0.1  \\
      Mg & 0.1  \\
      Si & 0.2  \\
      S  & 0.008  \\
      Ar & 0.0002 \\
      \hline
    \end{tabular}}\label{tab:first}
\begin{tabnote}
\end{tabnote}
\end{table}

If the total mass of the progenitor system is around $2.6~\Msun$ as in the case introduced in Section~\ref{sec:progenitor}, a bound remnant of around $2~\Msun$ remains with our assumption of $M_\mathrm{CSM}=0.3~\Msun$ and $M_\mathrm{ej}=0.3~\Msun$. The bound remnant can eventually experience core collapse induced by electron capture after $100-1000~\mathrm{years}$ \citep{wu2023b} and can be observed as a stripped-envelope electron-capture SN \citep{moriya2016}.

In summary, the light-curve properties of SN~2021yfj can be explained by the interaction between the ejecta with $E_\mathrm{ej}=4\times10^{50}~\mathrm{erg}$ and $M_\mathrm{ej}=0.3~\Msun$ and the dense CSM with $0.3~\Msun$ extended to $10^{15}~\mathrm{cm}$. As we performed an order of magnitude estimates, the estimated values have uncertainties of a factor of a couple. \citet{schulze2025} presented light-curve and spectral models assuming $M_\mathrm{ej}\simeq 3-5~\Msun$. However, as discussed so far, the light-curve properties can be explained by $M_\mathrm{ej}\simeq 0.3~\Msun$. If the estimated explosion and CSM properties are consistent with the observed spectra needs to be investigated, but the interaction-dominated spectra are likely reproduced as long as the strong interaction exists. Because the CSM is estimated to be optically thick, an aspherical CSM configuration may be required to explain the observed early spectra having P-Cygni profiles. The differences in the estimated properties originate from the degeneracies in the parameter estimation in interaction-powered SNe. 

Through spectral modeling of SN~2021yfj, \citet{schulze2025} estimated the elemental composition of the CSM in SN~2021yfj to be 78.6\% (O), 10\% (Ne), 5\% (Si), 3\% (S), 1\% (Ar), and 1\% (Mg). The average elemental abundances of the outer $0.3~\Msun$ of the hybrid WD, which is assumed to become the CSM after the DWD merger, are listed in Table~\ref{tab:first}. \citet{schulze2025} found that the Si and S abundances need to be of the order of 1\% to explain the Si and S features of SN~2021yfj. While the S fraction in the CSM in our model is around 1\%, the Si fraction is 20\%. Thus, our model may result in stronger Si emission than observed in SN~2021yfj although the detailed spectral modeling needs to be performed to confirm. In addition, we only present one example of the hybrid WD model in this paper. Further studies on the diversities in the hybrid WD compositions are required (Wu et al. in preparation). The existence of He, which is difficult to explain with the massive star progenitor scenario, is naturally explained in the DWD meger scenario. This is because He needs to be accreted from the He star companion to form the hybrid WD. He remains near the surface of the hybrid WD even after the mass transfer phase (Figure~\ref{fig:abundance}).

The CSM formed through the tidal stripping needs to be extended to $10^{15}~\mathrm{cm}$ to explain SN~2021yfj. 
Some numerical simulations of DWD mergers predict the existence of such an extended component in tidally stripped matter \citep[e.g.,][]{raskin2013,zetani2023}, while others do not \citep[e.g.,][]{pakmor2012,wu2023b}. The mass of the stripped matter is similarly uncertain.
However, SN~2020aeuh provided observational indications that the CSM formed through a DWD merger can be extended above $10^{15}~\mathrm{cm}$ and its mass can be as massive as $1~\Msun$ \citep{tsalapatas2025}. SN~2020aeuh showed signatures of the interaction between SN~Ia ejecta and a dense C+O CSM and thus it was interpreted to come from a merger of two C+O WDs forming the C+O CSM and triggering a double degenerate SN~Ia explosion. Some Ca-strong SNe related to DWD mergers are also observed to show signatures of extended CSM \citep{jacobson-galan2022}. SN~2021yfj can originate from a similar DWD merger with a hybrid WD with an outer Si and S-rich dense layer.

SN~2021yfj appeared in a star-forming galaxy \citep{schulze2025}. The delay time of DWD mergers from He stars is relatively small \citep{wang2017} and thus SN~2021yfj-like events from DWD mergers can also prefer to appear in star-forming galaxies. The event rate of SN~2021yfj-like events are estimated to be less than 0.1\% of Type~Ibc SNe \citep{schulze2025}. The event rate of the DWD mergers involving hybrid WDs must be rare, which is consistent with the very low event rate of SN~2021yfj-like events. A quantitative investigation of the event rate of SN~2021yjf-like events will be discussed in a forthcoming paper (D. Liu et al. in preparation).

\section{Discussion and conclusions}\label{sec:discussion}
We have demonstrated that Si and S-rich CSM can be formed through DWD mergers with hybrid WDs having outer Si and S-rich layers. Such DWD mergers can lead to SN~2021yfj-like events interacting with Si and S-rich CSM. While we investigated rather rare systems containing hybrid WDs, similar DWD mergers can occur with systems containing C+O WDs and He WDs. The resulting interacting SNe can be observed as SNe~Icn and SNe~Ibn, respectively. Thus, it is possible that some SNe~Ibn and Icn originate from DWD mergers \citep[e.g.,][]{wu2024}. In fact, although most SNe~Ibn and Icn are associated with star-forming regions indicating that their progenitors are massive stars, a small fraction originates from regions without any trace of ongoing star-forming activities \citep[e.g.,][]{sanders2013,hosseinzadeh2019,warwick2025,hu2026,dong2025}. These rare populations could be associated with DWD merger events. 

\citet{schulze2025} proposed to classify SN~2021yfj-like events as SNe~``Ien.'' This proposed classification is based on the interpretation that the progenitor of SN~2021yfj was a massive star. Motivated by the interpretation that the dense He CSM in SNe~Ibn is from the He layer of massive stars and the dense C+O CSM in SNe~Icn is from the C+O layer of massive stars, SNe~``Ien'' are named by assuming that the Si+S-rich CSM is from the Si+S-rich layer of massive stars. They reserved SNe~``Idn'' for SNe interacting with the O+Ne CSM originating from the O+Ne layer of massive stars. However, it is better for the observational classification scheme not to involve physical interpretations of SNe. The classifications of SNe~Ib and Ic are not based on the interpretation that they originate from massive stars forming He and C+O layers, but they are purely spectroscopic based on the existence and non-existence of He. In fact, SNe with prominent Si absorption features, containing S as well, are called SNe~Ia. In this respect, SNe with prominent narrow Si and S emission features should be called SNe~``Ian'', but this may be rather confusing and we may require another classification for SN~2021yfj-like events.

This paper presented that it is possible that DWD mergers can lead to SN~2021yfj-like events. Further studies such as detailed abundance computations and event rate estimates are required to see if this scenario is compatible with SN~2021yfj-like events. Further studies of the SN properties like numerical light-curve and spectral computations are also necessary. Discoveries of diverse interacting SNe may indicate they have diverse origins not limited to massive stars. DWD mergers could have much richer outcomes than previously considered.

\begin{ack}
We would like to thank the anonymous referee for constructive comments that improved this paper.
Numerical computations were in part carried out on PC cluster at the Center for Computational Astrophysics, National Astronomical Observatory of Japan.
\end{ack}

\section*{Funding}
TJM is supported by the Grants-in-Aid for Scientific Research of the Japan Society for the Promotion of Science (JP24K00682, JP24H01824, JP21H04997, JP24H00002, JP24H00027, JP24K00668) and by the Australian Research Council (ARC) through the ARC's Discovery Projects funding scheme (project DP240101786). 
CYW, DDL, ZWL and BW are supported by the National Natural Science Foundation of China (Nos 12288102, 12225304, 12473032, 12273105 and 12090040/2021YFA1600401/12090043), the National Key R\&D Program of China (Nos. 2021YFA1600403 and 2021YFA1600400), the Youth Innovation Promotion Association CAS (No. 2021058),the Yunnan Revitalization Talent Support Program (Young Talent Project, Yunling Scholar Project and Innovation Team Project), International Centre of Supernovae (ICESUN), Yunnan Key Laboratory of Supernova Research (No. 202505AV340004), and the Yunnan Fundamental Research Projects (Nos 202401AV070006, 202501AW070001, 202201BC070003, 202501AS070005 and 202401BC070007), the Strategic Priority Research Program of the Chinese Academy of Sciences (grant Nos. XDB1160303, XDB1160300, XDB1160000) and the Yunnan Revitalization Talent Support Program--Science \& Technology Champion Project (No.202305AB350003).  

\section*{Data availability} 
The data underlying this article will be shared on reasonable request to the corresponding author.



\bibliographystyle{apj}
\bibliography{pasj}

\end{document}